\documentclass[a4paper]{jpconf}
\usepackage{graphicx}
\usepackage{subfig}

\begin{document}
\title{Bus Stop’s Location and Bus Route Planning Using Mean Shift Clustering and Ant Colony in West Jakarta}

\author{Kenny Supangat}

\address{\textit{System Information Dept.}, Multimedia Nusantara University, Banten, Indonesia}
\ead{kenny.supangat@student.umn.ac.id}

\author{Yustinus Eko Soelistio}

\address{\textit{System Information Dept.}, Multimedia Nusantara University, Banten, Indonesia}

\ead{yustinus.eko@umn.ac.id}

\begin{abstract}
Traffic Jam has been a daily problem for people in Jakarta which is one of the busiest city in Indonesia up until now. Even though the official government has tried to reduce the impact of traffic issues by developing a new public transportation which takes up a lot of resources and time, it failed to diminish the problem. The actual concern to this problem actually lies in how people move between places in Jakarta where they always using their own vehicle like cars, and motorcycles that fill most of the street in Jakarta. Among much other public transportations that roams the street of Jakarta, Buses is believed to be an efficient transportation that can move many people at once. However, the location of the bus stop is now have moved to the middle of the main road, and it’s too far for the nearby residence to access to it. This paper proposes an optimal location of optimal bus stops in West Jakarta that is experimentally proven to have a maximal distance of 350 m. The optimal location is estimated by means of mean shift clustering method while the optimal routes are calculated using Ant Colony algorithm. The bus stops locations rate of error is 0.07\% with overall route area of 32 km. Based on our experiments, we believe our proposed bus stop plan can be an interesting alternative to reduce traffic congestion in West Jakarta.
\end{abstract}

\section{Introduction}
\hspace{5mm}According to the new start-stop index created by Castrol motor oil company [1], Jakarta has become number one city with the worst traffic in the world. The fact that Jakarta already has quite a variety of transportation like trains, angkot (public car established by private company), public bus (runs by a private company), ojek (a motorcycle runs by an individual person with low rate of fare), and the newest addition to it is the Busway (BRT-Bus Rapid Transit established by the Official Government of Jakarta) are still not proven to be an optimal solution for solving the traffic problem. Moreover, the actual issue that shows a huge contribution to the traffic problems was the increasing number of a private vehicle in Jakarta streets [2].The survey presented by the Statistic Department of Central Jakarta illustrate the number of a private two-wheel motorcycle is filling most of the roads with roughly 13.084.372 units or 74.67\% on the road. In contrast, other vehicles especially buses only take 2.07\% part of the population, while the other like private cars, trucks, buses, and governments vehicles are each takes the percentage of 18.64\%, 3.84\%, and 0.79\% respectively [2].

The private vehicle owners mostly absorb the total usage of public transportation, especially the private car and motorcycle owner. The bus is one of a few public transportation that get the least attention from people due to some reason presented in the early survey. The survey asked 46 respondents (24 male and 22 female) about the transportation they usually use with their own reason. The survey shows that 63\% (29 people) of the respondents don't mainly use bus because “the bus is too far away from my home”, the other reasons are “there are no bus stop nearby my residence” with 27\%, and just 10\% of the respondent felt awfully uncomfortable with the bus condition.

According to the data found during the survey, this research will be focusing on finding the optimal location for bus stops that will be located relatively near to the residential area, and the shortest route for the newly discovered bus stops location in this research.

\section{Data Acquisition}
\hspace{5mm}The residential area located in West Jakarta holds the largest number of houses in Jakarta region [3]. Moreover, the region includes several well-known residential districts in the center of West Jakarta area (i.g. Green Garden, Kembangan, Kedoya district, and more), along with more than five thousand houses outside these districts. A large portion of West Jakarta have been chosen as the designated area for this experiment, one of the areas is shown in Figure 1 where the houses serve as the main data is marked as red pins in the figure. 

\begin{figure}[]
\begin{center}
\includegraphics[scale=0.5]{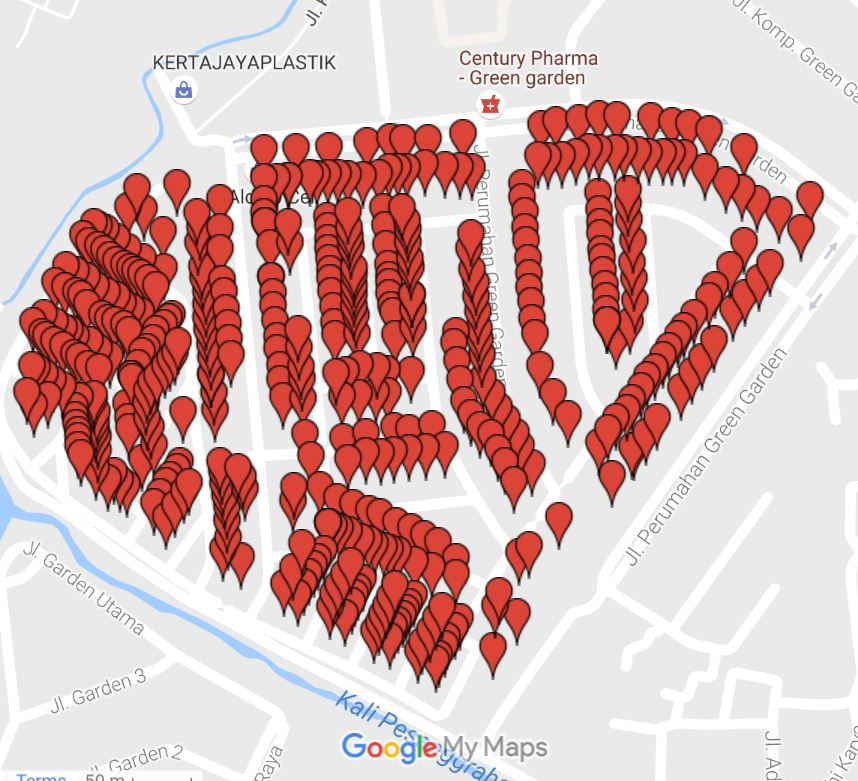}
\caption{\label{label}Example of part of West Jakarta on Google Maps.  Red marks are annotated houses.}
\end{center}
\end{figure}

\hspace{5mm}The file containing the whole data will later serve as data input to finding the optimal bus stop locations by using mean shift algorithm implemented in python version 2.7.11.

\section{Methodology}
\subsection{Bus Stop's Location Estimation}
\hspace{5mm}  We apply mean shift [4-6] [11] (Equation 1) clustering algorithm to estimate the optimal location of the bus stops.   The location of the bus stops are represented by the cluster center calculated by the mean shift, and the radius of the bus stop's service areas are the bandwidth of the clusters.

\begin{large}
\begin{equation} \label{eq:meanShift}
m(x)=\frac{\sum_{s\in S}K(s - x)s}{\sum_{s\in S}K(s - x)}
\end{equation}
\end{large}

Where the difference of m(x) - x is called as mean shift in [4], x is initial centroid estimates (the longitude and latitude of inputs), S is the longitude and longitude of labeled houses, and K is the kernel used in the method (Gaussian kernel with $K(s-x)=e^{-c||s-x||^2}$). We also using a bandwidth or radius that represent how far is the service area of the new bus stop will be, in this case, we will set an arbitrary range of bandwidth of 500m as the default bandwidth for the first attempt in this experiment. 

\subsection{Bus Stops Routes Optimization}

Previous studies have shown that ant colony can be used to solve optimal route problem (Bedi, 2007 [13]; Alves, 2010 [14]).The bus stops routes are optimized using ant colony algorithm [7] [8] [12] following:

\begin{large}
\begin{equation}
P_{ij}=\frac{(\tau_{ij}^\alpha)(\eta_{ij}^\beta)}{\sum(\tau_{ij}^\alpha)(\eta_{ij}^\beta)}
\end{equation}
\end{large}

where ,$P-ij$. is the probability of choosing the state, $T-ij$. is the intensity of the pheromone trail for each interstate, ,$N-ij$. is the visibility of a solution that would be selected by the ants. For the pheromone part, $\alpha$ is a parameter controlling the intensity of the pheromone trail where  $\alpha\geq0$, and $\beta$ is a parameter controlling the visibility where $\beta\geq0$ \cite{Haryanto}.

The new pheromone trail $(T-ij)$ is recalculated in each state by:

\begin{large}
\begin{equation}
\tau_{ij}=(1 - \rho)\tau_{ij} + \rho \Delta \tau_{ij} 
\end{equation}
\end{large}

\noindent where $\rho$ is the constant evaporation of pheromone trail $0 >\rho>1$ \cite{Haryanto}.

The ant colony algorithm is implemented in 2013 version of Matlab. In our implementation we use $\alpha=4$ and $\beta=1$ as the main parameter of both the intensity of the pheromone trail and ants visibility. To calculate the pheromone trail, we set the $\rho=0.15$.

\section{Result and Analysis}

\subsection{Bus Stops Location}

\begin{figure}[!tbp]
  \centering
  \subfloat[Mean shift result with 500m bandwidth.]{\includegraphics[width=0.45\textwidth]{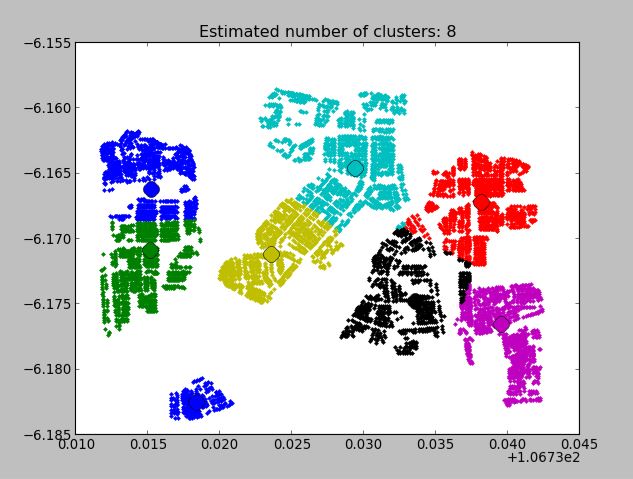}\label{fig:bandwidth500}}
  \hfill
  \subfloat[Mean shift result with 350m bandwidth.]{\includegraphics[width=0.45\textwidth]{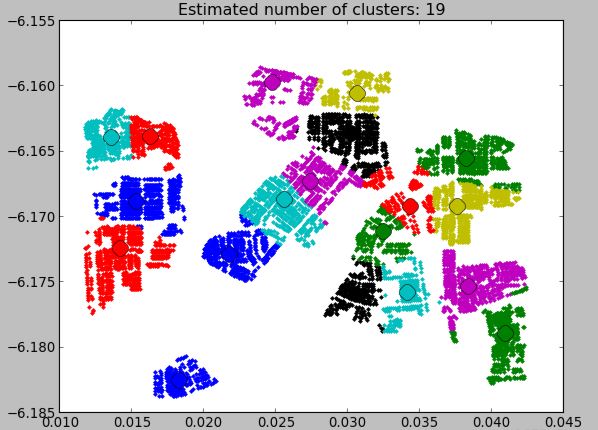}\label{fig:bandwidth350}}
  \caption{Mean shift results with two different bandwidths.  Notice that smaller bandwidth produces more cluster center (bus stops).}
\end{figure}

\hspace{5mm}We estimate the location of bus stops using an arbitrary 500 meters bandwidth on Equation \ref{eq:meanShift}.  It produces eight cluster centers that will be used as the new location for the new bus stop location (Figure \ref{fig:bandwidth500}. Unfortunately, the result is not quite reliable, because in the real situation every road in the street is full of curves and not just a straight road. 

To find the amount of error produced by the first result, a customized google maps API code is made to calculate how many houses miss the actual bus stop service radius or we can say further than 500m from house to bus stop. There are 594 houses or 7.72\% out of 7962 total houses that miss the bus stop service radius, and also the other error information is the maximum, minimum, and average range of error of each house.

After seeing the error result, we decide to use various numbers of arbitrary bandwidth radius ranging from 450m, 400m, 350m, 300m, and 250m. From each of the result that can be seen in Table \ref{table:meanShift}, and we choose one result that has the least error as the final solution.

\begin{table}[]
\begin{center}
\caption{\label{table:meanShift}Summary of mean shift result.  The best fit radius is 350 m indicated by the lowest error percentage.}
\begin{tabular}{| c | c | c | c | c | c | c |}
\hline
Bandwidth&500 m&450 m&400 m&350 m&300 m&250 m\\
\hline
Total Error&594&22&57&6&10&7\\
\hline
Error Percentages&7.72\%&0.28\%&0.74\%&0.07\%&0.13\%&0.09\%\\
\hline
Max Error&0.4 km&0.07 km&0.08 km&0.09 km&0.11 km&0.1 km\\
\hline
Min Error&0.01 km&0.01 km&0.01 km&0.01 km&0.01 km&0.01 km\\
\hline
Median Error&0.09 km&0.03 km&0.02 km&0.02 km&0.03 km&0.07 km\\
\hline
Bus Stops Spawned&9&14&14&19&22&33\\
\hline
\end{tabular}
\end{center}
\end{table}

From the result illustrated by the Table \ref{table:meanShift}, shows a various result from different bandwidth (radius). We pick the 350 m bandwidth because the result produces least error percentage with 0.07\%, i.g. six houses miss the boundary of 350 m.

From the result using 350 m bandwidth in Figure \ref{fig:bandwidth350}, we can see that there is 19 bus stop generated in the map, compare with Figure \ref{fig:bandwidth500} that use 500 m bandwidth which only generates eight bus stops with higher bandwidth radius. This means the result of 350 m bandwidth can reach out smaller area than the result from 500 m bandwidth.

\newpage

\begin{figure}[]
  \centering
  \subfloat[The old and the new bus stops location.  Red and blue squares represent the existing (old) and the proposed (new) bus stops location respectively.]{\includegraphics[width=0.45\textwidth]{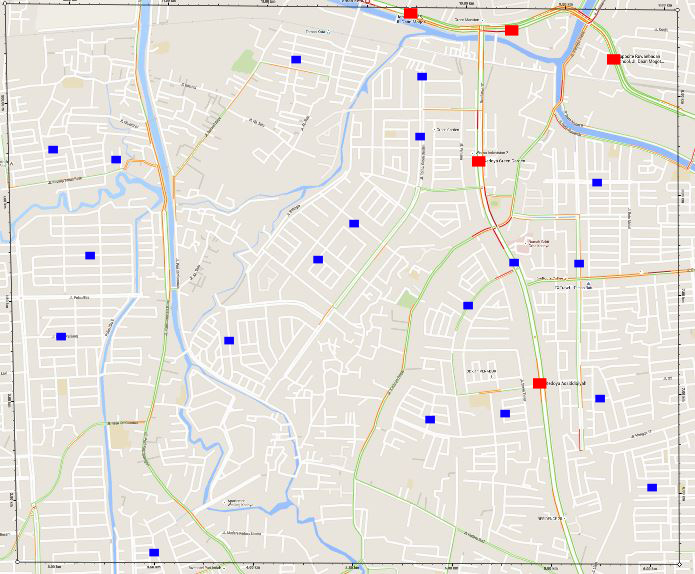}\label{fig:newButStops}}
  \hfill
  \subfloat[New proposed bus routes.  Red marks are the bus stops.  Blue paths indicate route traveled.  The numbers show the sequence of the stops.]{\includegraphics[width=0.40\textwidth]{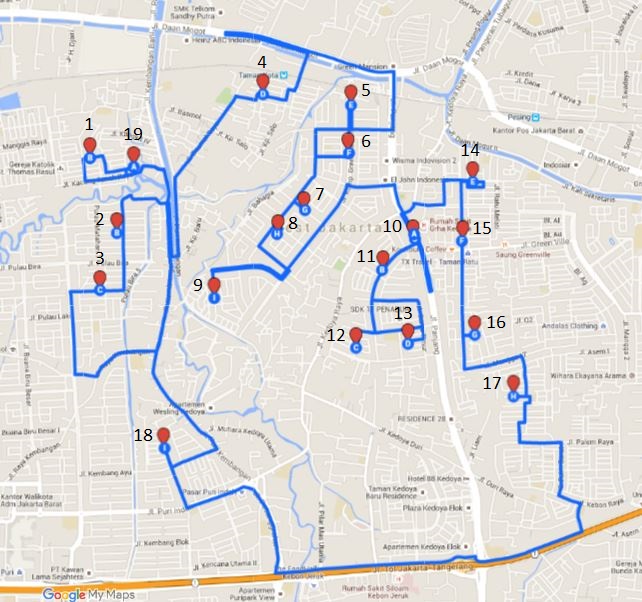}\label{fig:finalRoutes}}
  \caption{Mean shift results with two different bandwidths.  Notice that smaller bandwidth produces more cluster center (bus stops).}
\end{figure}

The placement of the new bus stop and the existing bus stops can be seen in Figure \ref{fig:newButStops}. The blue box represents the new bus stop, and the red box represents the old bus stop. We can see there are only 4 bus stops marked with a red box that currently operating and located in the middle of the main road far from the residential area . In the other hand, there are 19 new bus stops marked with a blue box scattered around the places so that many people can access the bus stop more easily across the residential area.

\subsection{Bus Route}

\hspace{5mm}From the chosen result before, the geographical location of the new bus stop locations will be inserted into the source code as the input by using Matlab programming tools. There are several main inputs used in the like the number of ants, a number of cities (in here will be served as bus stop geographical location) and a number of iteration or repetition because the ant colony works in random and has many different kinds of possible outputs \cite{Dorigo3}.


As presented in Figure \ref{fig:finalRoutes} are the final result of ant colony method where the route leads from bus stop from number 1 through 19 and back to bus stop number 1. The visualization of the route is created using Google Maps feature.

The 26.44 km route comes from several run attempts of changing the parameter of ant colony method. From Table 2, 3, 4, and 5 are the attempt to find the shortest by changing some parameter in ant colony method which is number of ants, and number loop or attempt that the ants going to do to find the shortest path in that attempt.

\section{Conclusion}

\hspace{5mm}This study has successfully determined the optimum solution for bus stop locations and its sub-optimal route. We prove that mean shift and ant colony algorithms can handle this particular problem, and should the case where the area is expanded further. And based on this experiments, we believe our proposed bus stop plan can be an interesting alternative to reduce traffic congestion in West Jakarta.  Next we should expand the area of interest to cover wider Jakarta area and calculate more optimize bandwidth to find best fit radius.


\section*{References}
\begin{thebibliography}{9}

\bibitem{Toppa}Toppa, Sabrina, (2015, February). These Cities Have The Worst Traffic in the World, Says a New Index. TIME, Retrieved from http://time.com/3695068/worst-cities-traffic-jams/?iid=sr-link1. 
\bibitem{Statistic}Statistic Department Center of DKI Jakarta Province, “Transportation Statistic of DKI Jakarta 2015”, DKI Jakarta BPS Province, 2015
\bibitem{StreetDirectory}StreetDirectory, "Komplek Perumahan in Jakarta Barat". Retrieved from http://www.streetdirectory.co.id/indonesia/jakarta/landmark/zone/jakarta+barat/komplek+perumahan/.
\bibitem{Fukunaga}Fukunaga, Keinosuke, and Larry D. Hostetler. "The estimation of the gradient of a density function, with applications in pattern recognition."Information Theory, IEEE Transactions on 21.1 (1975): 32-40.
\bibitem{Cheng}Cheng, Yizong. "Mean shift, mode seeking, and clustering." Pattern Analysis and Machine Intelligence, IEEE Transactions on 17.8 (1995): 790-799.
\bibitem{Cheng}Cheng, Yizong, and King-Sun Fu. "Conceptual clustering in knowledge organization." Pattern Analysis and Machine Intelligence, IEEE Transactions on 5 (1985): 592-598.
\bibitem{Dorigo}Dorigo, Marco, et al., eds. Ant Colony Optimization and Swarm Intelligence: 6th International Conference, ANTS 2008, Brussels, Belgium, September 22-24, 2008, Proceedings. Vol. 5217. Springer, 2008.
\bibitem{Dorigo2}Dorigo, Marco, and Thomas Stützle. "Ant colony optimization: overview and recent advances." Techreport, IRIDIA, Universite Libre de Bruxelles (2009).
\bibitem{Dorigo3}Dorigo, Marco, and Luca Maria Gambardella. "Ant colonies for the travelling salesman problem." BioSystems 43.2 (1997): 73-81.
\bibitem{Haryanto}Haryanto, Ardy Wibowo, Adhi Kusnadi, and Yustinus Eko Soelistio. "Warehouse Layout Method Based on Ant Colony and Backtracking Algorithm." arXiv preprint arXiv:1508.04872 (2015).
\bibitem{Comaniciu}Comaniciu, Dorin, and Peter Meer. "Mean shift: A robust approach toward feature space analysis." Pattern Analysis and Machine Intelligence, IEEE Transactions on 24.5 (2002): 603-619.
\bibitem{Dorigo4}Dorigo, Marco, and Luca Maria Gambardella. "Ant colony system: a cooperative learning approach to the traveling salesman problem."Evolutionary Computation, IEEE Transactions on 1.1 (1997): 53-66.
\bibitem{Bedi}Bedi, Punam, et al. "Avoiding traffic jam using ant colony optimization-a novel approach." Conference on Computational Intelligence and Multimedia Applications, 2007. International Conference on. Vol. 1. IEEE, 2007.
\bibitem{Alves}Alves, Diogo, et al. "Ant colony optimization for traffic dispersion routing." Intelligent Transportation Systems (ITSC), 2010 13th International IEEE Conference on. IEEE, 2010.

\end{thebibliography}
\end{document}